\documentclass[12pt]{article}
\usepackage{amssymb}
\usepackage{color}
\begin{document}

\begin{center}
   {\Large \textbf{Putnam Physics Competition}}\\
   {\large          \textbf{a suggestion}}
\end{center}

\vspace{5mm}
\begin{center}
   {\bfseries Costas J. Efthimiou}\\
   Department of Physics\\
   University of Central Florida\\
   Orlando, FL 32816, USA

   \vspace{1cm}
   \footnotesize
   dedicated to\\
   Oxylos and Ifitos, founders of \textit{The Olympic Games}\\
   W.L. Putnam, visionary of \textit{The Putnam Mathematical Competition}\\
   \normalsize
\end{center}

\vspace{1cm}
%%%%%%%%%%%%%%%%%%%%%%%%%%%%%%%%%%%%%%%%%%%%%%%%%%%%%%%%%%%%%%%%%%%%%%%%%%%%

``\emph{And we compel men to exercise their bodies not only for games,
  so they can win
  the prizes---for very few of them go to them---but to gain a greater
  good from it
  for the whole city, and for the men themselves}".
This is what Lucian of Anacharsis wrote ca. AD 170 for the victory
in \textit{The Olympic Games}. Being victorious in the games in
ancient Greece became a major achievement that gave credit not
only to the athlete but to his city-state as well. The athletes'
wide recognition for their physical abilities was (and still is)
the highest prize that made them to develop an  ardor, often an
obsession, for the competition. However, personal achievement
could not be imagined without the contribution, and therefore,
acknowledgement of the athlete's city-state.  Personal athletic
victory was identified with the victory of his city-state. The
city-state became the only representative and collective body with
the right to receive glory and awards. During \textit{The Olympic
Games}, all Greek cities-states could send official missions to
attend the games. There, famous poets and historians promoted
their works. And, famous (natural and other) philosophers
exchanged and debated ideas. These national gatherings promoted
cultural consciousness and strengthened the Greek identity.

L.W. Putnam recognized in undergraduate students the same virtue
which the ancient Greek athletes possessed \cite{Putnam}: ``The
idealism of the undergraduate student, his eagerness to achieve
something for his college, for his country or for any cause which
fills him with enthusiasm is constantly referred to with
admiration by those in charge of universities... In none of these
cases is the undergraduate primarily interested in winning honor
for himself. He is anxious ... and very glad to play a useful ...
part in the preparation of the team by which her victory is
secured." He thus proposed the establishment of a competition at
the college and university level: ``It seems probable that the
competition which has inspired young men to undertake and undergo
so much for the sake of athletic victories might accomplish some
result in academic fields." This vision was finally realized in
\textit{The Putnam Mathematical Competition}.

The mathematical community in North America is well-informed about
\textit{The Putnam Mathematical Competition} which ``has been a
factor of the outmost importance in arousing and stimulating
interest in mathematics in the colleges and universities of the
United States and Canada... The competition has undoubtedly played
no small part in raising the status, the level and standards of
mathematical education." \cite{Mordell}. The benefits of the
competition are hard to underestimate. The competition has
promoted mathematical consciousness to undergraduate students, it
has strengthened the cooperation between the colleges and
universities, and has served as an instrument to establish a
strong mathematical identity. The students' wide recognition for
their mathematical abilities is the highest prize that makes them
to develop an ardor, often an obsession, for the competition.
However, personal achievement cannot be imagined without the
contribution, and therefore, acknowledgement of the student's
university. Mathematical personal victory is identified with  the
victory of the college or university and the college or university
is the only representative and  collective body with the right to
receive glory and awards\footnote{This is even truer with the
International Mathematical Olympiad competitions. The results of
the competitions are announced as a ranking of teams (countries),
not ranking of individual students although medals are awarded to
students. Individual names are meant to and will be forgotten but
the winning countries will be, for ever, remembered.}.

When such an important competition has already been organized for
70 years in the area of mathematics, it is quite surprising that
this has not been extended to other fields, and, in particular,
physics. In Putnam's words: ``... it is a curious fact that no
effort has ever been made to organize contesting teams in regular
college studies. All rewards for scholarship are strictly
individual and are given in money, or in prizes or in honorable
mention. No opportunity is offered a student by diligence and high
marks in examinations to win or help in winning honor for his
college. All that is offered to him is the chance of personal
reward. Little appeal is made to high ideals or to unselfish
motives."

And although there are several local competitions along these
lines, I would  like to bring to the attention of the physics
community this failure to include such an important global
activity among the large number of other ongoing activities.

We already know from the list of winners \cite{Birkhoff} in
\textit{The Putnam Mathematical Competition} that physics students
value the competition highly. The list includes many students who
became physicists and had a significant impact in the progress of
physics: R.P. Feynman (1939), R.L. Mills (1948), A. Zemach (1949),
P.J. Redmond (1951), T.T. Wu (1953), J.D. Bjorken (1954), K.G.
Wilson (1954, 1956), R.M. Friedberg (1956), S.L. Adler (1959).
Wouldn't these people have been even more enthusiastic to compete
in a physics competition? I know from personal experience that
this would be the case. I was fortunate enough to have won a prize
in a national competition and participate subsequently in the XXIV
International Mathematics Olympiad. However, I have always felt
sorry that I had never had the chance to compete in a Physics
Olympiad.
%as his country was not participating
%in the competition then.

As in the case of mathematics, the benefits of  a \emph{Physics
Competition} are hard to  underestimate. The competition will
promote scientific consciousness to undergraduate students, it
will strengthen furthermore the cooperation between the colleges
and universities, and it will serve as an instrument to establish
strong physics identity. The students' wide recognition for their
scientific abilities will make them to develop an ardor, perhaps
even an obsession, for the competition and physics which, in turn,
will serve as a means to increase the number of physics students
with extraordinary talent in a time when such an outcome is highly
desirable.

%%%%%%%%%%%%%%%%%%%%%%%%%%%%%%%%%%%%%%%%%%%%%%%%%%%%%%%%%%%%%%%%%%%%%%%%%%%%
%\newpage
\vspace{2cm}

 {\bfseries Appendix: A Possible Description of the
Competition}\\[5mm]

Establishing a syllabus that would be fair for all colleges and
universities is not an easy task. The issue has already been faced
in the case of the \textit{Putnam Mathematical Competition} where
various opinions have been expressed. The distinguished
mathematician L.J. Mordell has analyzed the problem from his
personal point of view in \cite{Mordell}. His article was replied
to by L.M. Kelly \cite{Kelly} who also expressed his personal
position and adopted a different perspective. Both articles are
useful resources toward the adoption of a syllabus, as well as a
refinement of the details of the competition.

Our suggestion is to allow the syllabus of the 22nd \textit{Putnam
Mathematical Competition} to become the guide of a possible
\textit{Putnam Theoretical Physics Competition} with the
appropriate adjustments in language to accommodate the `physics'
content of the competition:

\begin{center}
\begin{minipage}{12cm}
\small The examination will be constructed to test originality as
well as technical competence. It is expected that the contestant
will be familiar with the core courses embodied in undergraduate
physics (Newtonian Physics, Relativity, Electricity and Magnetism,
Thermodynamics, Statistical Physics, Quantum Mechanics). It is
assumed that such training, designed for physical science majors,
will include somewhat more sophisticated physical concepts than
introductory physics. Thus, topics such as calculus of variations,
linear algebra,  etc. and subtleties beyond the routine solution
devices are to be assumed. Questions will be included which cut
across the bounds of various fields and self-contained questions
which belong to more advanced areas may be included. It will be
assumed that the contestant has acquired a familiarity with the
body of physical lore and mathematics commonly discussed in
physics clubs or in courses with titles such as \textsf{Seminar on
the Foundations of Quantum Mechanics}. It is also expected that
self-contained questions involving group theory, atomic and
nuclear physics, etc. will not be entirely foreign to the
contestant's experience. \normalsize
\end{minipage}
\end{center}

The foregoing specifies a syllabus for a competition in
theoretical physics. A separate syllabus, and perhaps a separate
competition, may be established in experimental physics.
%However, the author does not feel confident enough to make
%statements for an experimental curriculum.

%%%%%%%%%%%%%%%%%%%%%%%%%%%%%%%%%%%%%%%%%%%%%%%%%%%%%%%%%%%%%%%%%%
\vspace{1cm}
\textbf{Acknowledgements}\\

I am thankful to Tristan H\"ubsch who read my original draft,
corrected many typos and commented on almost each and every of my
sentences. I have resisted to make all the changes he suggested so
the text faithfully represents my own beliefs and eccentricities.
If a handful of people read this manuscript as carefully and
passionately as he did, then I will have succeeded in my goal to
build an initial momentum towards the implementation of a national
college \emph{Physics Competition}.

\end{document}